\documentclass[11pt]{article}

\usepackage{amsmath}
\usepackage{amssymb}
\usepackage{amsfonts}

\def\av#1{\langle#1\rangle}
\def\d{\text{d}}

\def\df#1#2{\frac{\d#1}{\d#2}}

\def\d{\text{d}}

\def\vec#1{\mathbf{#1}}
\def\ave#1{\langle#1\rangle}
\def\tp{t_\text{P}}
\def\mp{m_\text{P}}

\begin{document}

\title{Probing spacetime fluctuations using cold atom traps}


\author{
Charles H.-T. Wang\\
SUPA Department of Physics, University of Aberdeen,
Aberdeen AB24 3UE, UK\\
\\
Robert Bingham\\
STFC Rutherford Appleton Laboratory, Chilton, Didcot, Oxfordshire OX11 0QX, UK\\
\\
J. Tito Mendon\c{c}a\\
IPFN, Instituto Superior T\'ecnico, 1049-001 Lisboa, Portugal
}

\date{}

\maketitle

\begin{abstract}
We investigate the position oscillation of a particle that models the center of mass quantum state of a trapped Bose-Einstein condensate coupled to the zero-point fluctuations of the gravitational field. A semiclassical analysis is performed that allows to calculate the mean square amplitude of the oscillation. In analogy with the Lamb shift in quantum electrodynamics, this gives rise to an upshift of the energy of the trapped condensates. We show that for an elongated trap, the energy shift scales quadratically with the length as well as cubically with the total number of atoms.
\end{abstract}

\section{Introduction}

The unprecedented precision of atom interferometry holds out new prospects for laboratory tests of general relativity \cite{Dimopoulos2007}. The development of large molecule interferometry is providing an arena to challenge our understanding of the foundations of quantum mechanics \cite{Hackermueller2004}. Furthermore, there has been a surge of interest in detecting signatures of low energy quantum gravity and unified theories using quantum matter waves \cite{HYPER2004, GAUGE2009, Amelino2009}.

The lowest energy quantum gravity effect is the zero-point vacuum fluctuations of spacetime. A possible scenario is the decoherence of matter waves through coupling to a fluctuating metric \cite{Percival2000, Wang2006, Bonifacio2009, Reynaud2006}. Such efforts can however be limited by the smallness of the effect and hence high sensitivity requirement \cite{Reynaud2004}, the need to suppress other environmental decoherence effects within the interferometer \cite{Hackermueller2004}, and the ambiguity of predictions in the absence of a consistent quantum theory of gravity \cite{Anastopoulos2008}.

One important consequence of the vacuum fluctuations of a quantum field is the energy shift of a bound state coupled to the field. The measurement of the energy or corresponding frequency shift through spectroscopy or resonance techniques can generally be done more accurately than the visibility measurement for quantum decoherence. Indeed, the resulting Lamb shift of electron energy levels in an atom provides the first experimental evidence for EM vacuum fluctuations and agrees with predictions from the renormalization procedures in QED.

In this paper we investigate the energy shift of quantum bound states due to spacetime fluctuations.  We have in mind a large number of trapped cold atom and but trapped molecules may also be relevant. For such a system, the energy shift is most important for its center-of-mass wavefunction. This motivates us to investigate the position oscillation of a particle that models the center of mass quantum state of a trapped Bose-Einstein condensate coupled to the zero-point fluctuations of the gravitational field. A semiclassical analysis is performed that allows to calculate the mean square amplitude of the oscillation. In analogy with the Lamb shift in quantum electrodynamics, this gives rise to an upshift of the energy of the trapped condensates. We show that for an elongated trap, the energy shift scales quadratically with the length as well as cubically with the total number of atoms.


\section{Fluctuations of gravitational field}

We adopt the linearized metric
\begin{eqnarray}\label{ggam}
g_{\mu\nu}
&=&
\eta_{\mu\nu}+ h_{\mu\nu}
\end{eqnarray}
about the Minkowski metric $\text{diag}(-1,1,1,1)$ with the metric perturbation $ h_{\mu\nu}$ in the transverse-traceless (TT) gauge:
\begin{eqnarray}
 h_{00} =  h_{0i}
&=&
0
\label{TT1}
\\
 h^i{}_i
&=&
0
\label{TT2}
\\
 h^j{}_{i,j}
&=&
0
\label{TT3}
\end{eqnarray}
where $i,j=1,2,3$.

In this gauge, for each wave vector $\vec{k}$ there are two polarisation tensors \cite{Flanagan2005}
\begin{eqnarray}
h_{i j }(\vec{k},1)
&=&
e_{i }(\vec{k},1)e_{j }(\vec{k},1)-e_{i }(\vec{k},2)e_{j }(\vec{k},2)
\label{polt1}
\\
h_{i j }(\vec{k},2)
&=&
e_{i }(\vec{k},1)e_{j }(\vec{k},2)+e_{i }(\vec{k},2)e_{j }(\vec{k},1)
\label{polt1a}
\end{eqnarray}
in terms of the two polarisation vectors $\vec{e}(\vec{k},\lambda)$ satisfying
\begin{eqnarray}
\sum_{\lambda=1}^{2}
e_{i}(\vec{k},\lambda)e_{j}(\vec{k},\lambda)
&=&
\delta_{ij}-k_{i}k_{j}/k^2 .
\label{pid2}
\end{eqnarray}

It follows from \eqref{pid2} that
\begin{eqnarray}
&&\hspace{-15pt}
\sum_{i,j=1}^{3}
\sum_{\lambda=1}^{2}
h_{i j }(\vec{k},\lambda)h_{i j}(\vec{k},\lambda)
=
\left[
\sum_{i=1}^{3}\sum_{\lambda=1}^{2}
e_{i}(\vec{k},\lambda)e_{i}(\vec{k},\lambda)
\right]^2
=
4 .
\label{poltid1}
\end{eqnarray}

We shall consider metric fluctuations at temperature $T$ using the following directional decomposition
\begin{eqnarray}
h_{i j }(\vec{r},t)
&=&
\sum_{\lambda=1}^{2}
\int\d^3 k \,h_{i j }(\vec{k},\lambda)F_T(\omega)
\cos[\vec{k}\cdot\vec{r}-\omega t - \theta(\vec{k},\lambda)]
\label{gwi}
\end{eqnarray}
in terms of a spectral density $F_T(\omega)$ and random angles $\theta(\vec{k},\lambda)$.
This gives rise to the following correlation relation:
\begin{eqnarray}
\ave{ h_{i j }(0,t) h_{l m }(0,t')}
&=&
\frac12
\sum_{\lambda=1}^{2}
\int\d^3 k \,
h_{i j }(\vec{k},\lambda)h_{l m }(\vec{k},\lambda)
F_T(\omega)^2
\cos\omega (t - t') .
\label{avegg}
\end{eqnarray}
Therefore using \eqref{poltid1}, Eq. \eqref{avegg} implies
\begin{eqnarray}
\ave{ h_{i j }(0,t) h_{i j }(0,t')}
&=&
\frac{8\pi}{c^3}
\int_0^\infty \omega^2
F_T(\omega)^2
\cos\omega (t - t')
\,\d\omega
\label{avegga}
\end{eqnarray}
For the equal time correlation, the above becomes
\begin{eqnarray}
\ave{ h_{i j } h_{i j }}
&=&
\frac{8\pi}{c^3}
\int_0^\infty \omega^2
F_T(\omega)^2
\,\d\omega .
\label{avegga0}
\end{eqnarray}

The function $F_T(\omega)$ can be fixed by comparing \eqref{avegga0} with the
the fluctuations of gravitational field at temperature $T$ satisfying \cite{Schafer1981}:
\begin{eqnarray}
\ave{ h_{i j} h_{ij}}
&=&
\frac{32}\pi\frac{G}{c^5}\int_0^\infty E_T(\omega)\d\omega
\label{avhh0}
\end{eqnarray}
where
\begin{eqnarray}
E_T(\omega)
&=&
\frac12\hbar\omega + \frac{\hbar\omega}{e^{\hbar\omega/kT}-1}
\label{Eome}
\end{eqnarray}
is the the Planck spectral density. This expression ensures that
the energy density of the fluctuating gravitational field is given by
\begin{eqnarray}
u
&=&
\frac1{\pi^2c^3}\int_0^\infty\omega^2E(\omega)\d\omega
\label{u}
\end{eqnarray}
having the same expression for the EM energy density at temperature $T$ \cite{Callen1951}.
Comparing \eqref{avegga0} and \eqref{avhh0} we see that in particular at zero temperature $T=0$:
\begin{eqnarray}
F_T(\omega)^2
&=&
\frac{2}{\pi^2}\,\frac{G\hbar}{c^2}\, \frac1\omega .
\label{FT}
\end{eqnarray}

Using \eqref{FT}, we see that \eqref{avegga} becomes
\begin{eqnarray}
\ave{ h_{i j }(0,t) h_{i j }(0,t')}
&=&
\frac{16}{\pi}\,\tp^2
\int_0^\infty
\omega
\cos\omega (t - t')
\,\d\omega
\label{aveggax}
\end{eqnarray}
and \eqref{avegga0} becomes
\begin{eqnarray}
\ave{ h_{i j } h_{i j }}
&=&
\frac{16}{\pi}\,\tp^2
\int_0^\infty \omega
\,\d\omega
\label{avegga0x}
\end{eqnarray}
where
\begin{eqnarray}
\tp
&=&
\sqrt{\frac{G\hbar}{c^5}}
\label{tp}
\end{eqnarray}
is the Planck time.

It then follows from \eqref{aveggax} that
\begin{eqnarray}
\ave{\dot{ h}_{i j}(0,t)\dot{ h}_{i j}(0,t')}
&=&
\frac{16}{\pi}\,\tp^2
\int_0^\infty
\omega^3
\cos\omega (t - t')
\,\d\omega
\label{aveggax0}
\end{eqnarray}
and
\begin{eqnarray}
\ave{ h_{i j, k}(0,t) h_{i j, k}(0,t')}
&=&
\frac{16}{\pi}\,\frac{\tp^2}{c^2}
\int_0^\infty
\omega^3
\cos\omega (t - t')
\,\d\omega .
\label{aveggax1}
\end{eqnarray}

The above infinite integrals need to be regularized with a UV cut-off angular frequency $\Omega$.
Using the identities
\begin{eqnarray}
\int_0^t\int_0^t
\cos\omega(t'-t'')\,\d t'\d t''
&=&
\frac{2}{\omega^2}\,(1-\cos\omega t)
\label{iint}
\end{eqnarray}
and
\begin{eqnarray}
\int_0^\Omega
\omega(1-\cos\omega t)\,\d\omega
&=&
\frac12\Omega^2
\left(
1-\frac{\sin\Omega t}{\Omega t}+\frac{1-\cos\Omega t}{\Omega^2 t^2}
\right) .
\label{intO}
\end{eqnarray}
We see that for a large time $t$ so that
$\Omega t \gg 1$, Eq. \eqref{intO} reduces to
\begin{eqnarray}
\int_0^\Omega
\omega(1-\cos\omega t)\,\d\omega
&=&
\frac12\Omega^2
\label{intO1}
\end{eqnarray}
In this limit, using \eqref{iint} and \eqref{intO1}, we further see that \eqref{aveggax0} becomes
\begin{eqnarray}
\int_0^t\int_0^t
\ave{\dot{ h}_{i j}(0,t')\dot{ h}_{i j}(0,t'')}
\,\d t'\d t''
&=&
\frac{16}{\pi}\,\tp^2\,\Omega^2 .
\label{Gam1}
\end{eqnarray}

\section{Geodesic perturbations}

To investigate the Brownian motion of a particle in the fluctuating space-time we shall consider the
geodesic equation
\begin{eqnarray}
\df{^2x^\rho}{\tau^2}
+
\Gamma^\rho{}_{\mu\nu}\df{x^\mu}{\tau}\df{x^\nu}{\tau}
&=&
0
\label{geod}
\end{eqnarray}
in the weak field limit
\begin{eqnarray}
\Gamma^\rho{}_{\mu\nu}
&=&
\frac12\eta^{\rho\lambda}{}( h_{\lambda\mu,\nu}+ h_{\lambda\nu,\mu}- h_{\mu\nu,\lambda})
\label{connex1}
\end{eqnarray}
where $\tau$ is the proper time.
The particle being considered is non-relativistic and so to the first order, the proper time can approximated using the coordinate time $\tau = ct$, and accordingly, $\df{}{\tau}$ becomes approximately $\frac1c\df{}{t}$.

We then consider a perturbed path due to the metric perturbation of the form
\begin{eqnarray}
x^\rho
&=&
x^{(0)}{}^\rho+x^{(1)}{}^\rho
\label{xrho}
\end{eqnarray}
where $x^{(0)}{}^\rho$ is the unperturbed path and $x^{(1)}{}^\rho$ is the perturbed displacement.

We further denote the corresponding time derivatives as
$u^\rho := \df{x^\rho}{t} = \dot{x}^\rho$
$u^{(0)}{}^\rho := \dot{x}^{(0)}{}^\rho$
$u^{(1)}{}^\rho := \dot{x}^{(1)}{}^\rho$
so that the perturbed velocity reads
\begin{eqnarray}
u^\rho
&=&
u^{(0)}{}^\rho+u^{(1)}{}^\rho
\label{urho}
\end{eqnarray}

Substituting \eqref{xrho} into \eqref{geod} we see that the unperturbed path simplify satisfies
\begin{eqnarray}
\ddot{x}^{(0)}{}^\rho
&=&
0
\label{ddrho}
\end{eqnarray}
i.e.
\begin{eqnarray}
\dot{u}^{(0)}{}^\rho
&=&
0
\label{du0}
\end{eqnarray}
implying the unperturbed velocity $u^{(0)}{}^\rho$ is constant.

The perturbed path then satisfies
\begin{eqnarray}
\ddot{x}^{(1)}{}^\rho
&=&
-\Gamma^\rho{}_{\mu\nu}\dot{x}^{(0)}{}^\mu\dot{x}^{(0)}{}^\nu
\label{geod3}
\end{eqnarray}
i.e.
\begin{eqnarray}
\dot{x}^{(1)}{}^\rho
&=&
{u}^{(1)}{}^\rho
\label{geod41}
\\
\dot{u}^{(1)}{}^\rho
&=&
-\Gamma^\rho{}_{\mu\nu}u^{(0)}{}^\mu u^{(0)}{}^\nu .
\label{geod42}
\end{eqnarray}






To illustrate the effect we consider a non-relativistic particle whose unperturbed motion is along the $x$-axis $(x=x^1)$ with a speed $v \ll c$ with $u^{(0)}{}^\mu = (c,v,0,0)$.
From \eqref{TT1}--\eqref{TT3} and \eqref{connex1} we see that
\begin{eqnarray}
\Gamma_{00\mu}
=
\Gamma_{\mu00}
&=&
0
\label{connex2}
\end{eqnarray}
\begin{eqnarray}
\Gamma^0{}_{0\mu}
=
\Gamma^{\mu}{}_{00}
&=&
0
\label{connex3}
\end{eqnarray}
and
\begin{eqnarray}
\Gamma^i{}_{0j}
&=&
\frac12\, h_{ij,0} .
\label{connex4}
\end{eqnarray}

Using \eqref{connex2} and \eqref{connex3}, and $v\ll c$ while keeping leading term proportional to $v$ but neglect $v^2$ terms, \eqref{geod41} and \eqref{geod42} become the following:
For $\rho=0$ we have
\begin{eqnarray}
\dot{x}^{(1)}{}^0
&=&
{u}^{(1)}{}^0
\label{geod51}
\\
\dot{u}^{(1)}{}^0
&\approx&
-\Gamma^0{}_{11}u^{(0)}{}^1 u^{(0)}{}^1
=
-\Gamma^0{}_{11}v^2
\nonumber
\\[3pt]
&\approx&
0 .
\label{geod520}
\end{eqnarray}
For $\rho=i=1,2,3$ we have
\begin{eqnarray}
\dot{x}^{(1)}{}^i
&=&
{u}^{(1)}{}^i
\label{geod52}
\\
\dot{u}^{(1)}{}^i
&\approx&
-2\Gamma^i{}_{01}u^{(0)}{}^0 u^{(0)}{}^1
=
-2\Gamma^i{}_{01}cv
\nonumber
\\[3pt]
&=&
- h_{1i,0}\,cv
=
-\dot{ h}_{1i}\,v
\label{geod521}
\end{eqnarray}
i.e.
\begin{eqnarray}
\ddot{x}^{(1)}{}^i
&=&
-\dot{ h}_{1i}\,v .
\label{geod522}
\end{eqnarray}

\subsection{Estimate of energy shift}

The correlation function of $\dot{ h}_{1i}$ will be estimated from \eqref{aveggax0} averaged over all metric components to be
\begin{eqnarray}
\ave{\dot{ h}_{1i}(t)\dot{ h}_{1i}(t')}
&=&
\frac{16}{9\pi}\,\tp^2
\int_0^\Omega
\omega^3
\cos\omega (t - t')
\,\d\omega
\label{g1i}
\end{eqnarray}

The approach here is analogous to the classical Brownian motion in a non-stationary configuration using the correlation assumptions in \eqref{avAx} and \eqref{avAu}. The initial position and velocity perturbations will be assumed to be zero:
\begin{eqnarray}
{x}^{(1)}{}^\rho(0)
&=&
0
\label{inix}
\\
{u}^{(1)}{}^\rho(0)
&=&
0 .
\label{iniu}
\end{eqnarray}
Then \eqref{u2int1}, \eqref{Gam1}, \eqref{g1i} give
\begin{eqnarray}
\av{u^{(1)}{}^i(t)^2}
&=&
v^2
\int_0^{t}
\int_0^{t}
\ave{\dot{ h}_{1i}(t')\dot{ h}_{1i}(t'')}
\,\d t'\d t''
\nonumber
\\[3pt]
&=&
\frac{16}{9\pi}\,\tp^2\,\Omega^2 v^2 .
\label{u2intx}
\end{eqnarray}
It follows from \eqref{avxdu} in the appendix that
\begin{eqnarray}
\df{^2\av{x^{(1)}{}^i(t)^2}}{t^2}
&=&
2\av{u^{(1)}{}^i(t)^2}
=
\frac{32}{9\pi}\,v^2\tp^2\,\Omega^2 .
\label{avxdux}
\end{eqnarray}
The solution of \eqref{avxdux} is given by
\begin{eqnarray}
\av{x^{(1)}{}^i(t)^2}
&=&
\frac{16}{\pi}\,\tp^2\,\Omega^2\,v^2t^2
=
\frac{16}{9\pi}\,\tp^2\,\Omega^2\ell^2
\label{avxduxsol}
\end{eqnarray}
where $\ell=vt$ is the unperturbed travel distance of the particle during time $t$.

Based on \eqref{avxduxsol} we now estimate the energy shift of a quantum particle of mass $m$ trapped in a cylinder of length $L$ subject to a transverse harmonic potential with angular frequency $\omega$.
For a particle having a mean free path length $\ell$ due to non-gravitational interactions along the axis of the cylinder, the mean square deviation from the axis follows as
\begin{eqnarray}
\av{\Delta \vec{r}^2}
&=&
\frac{16}{9\pi}\,\tp^2\,\Omega^2\ell^2
\end{eqnarray}
which induces a shift of the mean potential energy for the trap given by
\begin{eqnarray}
\av{\Delta V}
&\approx&
\frac{16}{27\pi}\,
{m \omega^2}\tp^2\,\Omega^2\ell^2 .
\end{eqnarray}
We further estimate that the mean free path length $\ell$ for the particle to be the length of the cylindrical trap $L$ and adopt the Compton frequency as cutoff:
\begin{eqnarray}
\Omega
&=&
\frac{m c^2}{\hbar} .
\end{eqnarray}
This leads to the estimated energy shift given by
\begin{eqnarray}
\av{\Delta V}
&\approx&
\frac{16}{27\pi}\,
\frac{m^3}{\mp^2} \omega^2\,L^2
\end{eqnarray}
where
\begin{eqnarray}
\mp
&=&
\sqrt{\frac{\hbar c}{G}}
\label{tp}
\end{eqnarray}
is the Planck time.

\section{Concluding remarks}

For example, assuming if $\omega=2\pi$ kHz and $N$ is the number of trapped rubidium atoms, then the ratio of the energy shift to the difference of transverse energy levels is
\begin{eqnarray}
\frac{\Delta E}{\hbar\omega}
\approx
4\times10^{-23} N^3 (L/\text{m})^2.
\label{ratio1}
\end{eqnarray}

For $N=10^6$ and $L=1$ cm, the gravitational Lamb shift is vanishingly small, with $\Delta E/\hbar\omega \approx 4\times10^{-9}$. With a relatively moderate improvement of atom number to $N=10^8$ while keeping $L=1$ cm, the energy shift yields $\Delta E/\hbar\omega \approx 0.5 \%$, a potentially measurable value. The discussion presented here is readily extended to other trap geometries. We expect that such traps will be available in the near future and the proposed experiment could potentially lead to an observable effect of low energy quantum gravity.

\section*{Acknowledgments}
The authors are grateful to the EPSRC and STFC Centre for Fundamental Physics for financial support.

\section*{Appendix: A simple Langevin equation}

Consider here a non-relativistic particle with position $x(t)$ and velocity $u(t)$ subject to a stochastic force $f(t)$ having a zero mean:
\begin{eqnarray}
\av{f}
&=&
0 .
\label{avA}
\end{eqnarray}
The governing equations are:
\begin{eqnarray}
\dot{x}&=&u
\label{dxu}
\\
\dot{u}&=&f .
\label{du}
\end{eqnarray}
The force is assumed to be uncorrelated with past position and velocity so that
\begin{eqnarray}
\av{f(t)x(t')}
&&
\left\{
  \begin{array}{ll}
    = 0, & t>t' \\[3pt]
    \neq 0, & t<t'
  \end{array}
\right. ;
\label{avAx}
\end{eqnarray}
and
\begin{eqnarray}
\av{f(t)u(t')}
&&
\left\{
  \begin{array}{ll}
    = 0, & t>t' \\[3pt]
    \neq 0, & t<t'
  \end{array}
\right. .
\label{avAu}
\end{eqnarray}
The particle's velocity can be readily integrated to be
\begin{eqnarray}
u(t)
&=&
u(0)+\int_0^{t}
f(t')\,\d t' .
\label{uint}
\end{eqnarray}
Averaging \eqref{uint} and using \eqref{avA} we have
\begin{eqnarray}
\av{u(t)}
&=&
u(0) .
\label{avu}
\end{eqnarray}
From \eqref{uint} we see that
\begin{eqnarray}
u(t)^2
&=&
u(0)^2
+
2u(0)
\int_0^{t}
f(t')\,\d t'
+
\int_0^{t}
\int_0^{t}
f(t_1)f(t_2)\,\d t_1\d t_2 .
\label{u2int}
\end{eqnarray}
Averaging \eqref{u2int} and using \eqref{avAu} we then have
\begin{eqnarray}
\av{u(t)^2}
&=&
u(0)^2
+
\int_0^{t}
\int_0^{t}
C_f(t'-t'')\,\d t'\d t''
\label{u2int1}
\end{eqnarray}
in terms of the correlation function
\begin{eqnarray}
C_f(t'-t'')
&=&
\av{f(t')f(t'')} .
\label{corrA}
\end{eqnarray}
Multiplying \eqref{du} by $x$ and using \eqref{dxu} we see that
\begin{eqnarray}
\df{^2(x^2)}{t^2}
&=&
2u^2+2x f .
\label{xdu}
\end{eqnarray}
On taking average and using \eqref{avAx}, this yields
\begin{eqnarray}
\df{^2\av{x^2}}{t^2}
&=&
2\av{u^2} .
\label{avxdu}
\end{eqnarray}
If $C_f(t)$ has a narrow peak near $t=0$ then \eqref{u2int1} can be approximated by
\begin{eqnarray}
\av{u(t)^2}
&=&
u(0)^2
+
\widetilde{C}_f(0)\,t
\label{u2intw}
\end{eqnarray}
in terms of the Fourier transform of $C_f(t)$
\begin{eqnarray}
\widetilde{C}_f(\omega)
&=&
\int_{-\infty}^{+\infty} C_f(t)\, e^{-i\omega t}\,\d t .
\label{FcorrA}
\end{eqnarray}

\end{document}